\documentstyle{article}
\begin{document}
\begin{center}
{\Large \bf One kind of spinor techniques for massive fermions}\\
V. V. Andreev
\footnote{E-mail: andreev@gsu.unibel.by} \\
Gomel State University, Phys.Department, Sovietskya str. 102, Gomel,
246699\\
Belarus \\
\end{center}

\begin{abstract}
\noindent
In the work we have considered the
reduction techniques  matrix element of the
reaction interaction of elementary particles
for massive fermions to spinor products .
The our procedure is not more complex than CALCUL spinor
techniques for massless fermions. It is relatively simple
and leads to basic spinor products with any polarizationod fermions.
Also obtained "spinor" Chisholm identities for massive fermions.
\end{abstract}

The study of the high-energy processes with polarization are of fundamental
importance in understanding structure of matter. For example, The European
Muon Collaboration (EMC) and SLAC experiments ~\cite{sp1} on deep
inelastic scattering of longitudinally polarized muons on
longitudinally polarized target has catalyzed an extraordinary
outburst of theoretical activity.

With increasing energy of colliders, processes involving many final-state
particles ( $2\to 3$, $2\to 4$, $2\to 5$, $\ldots$ ) are important path the
present collider's physics.

The above-listed directions of high-energy physics have generic point. The
calculation of cross sections for these processes is made difficult if you
used conventional approach. This approach is to square the Feynman amplitude
and because of it is very inconvenient to implement the number of Feynman
diagrams and the number of final state particles are large.

An alternative approach is to compute the Feynman amplitudes symbolically or
numerically. The idea of calculating Feynman amplitudes is as old as
conventional approach. For example, covariant method of calculating
amplitudes have developed more thirty years ago
~\cite{sp2}-\cite{sp4}.

Many different methods exist now.A particularly convened techniques for massless external fermions was
introduced by the CALCUL collaboration ~\cite{sp5}-\cite{sp7}.
Although, there have been several attempts to generalize the method
do the CALCUL collaboration for massless fermions to massive case,
the resulting procedure is rather cumbersome and requires the
introduction of arbitrary reference momenta~\cite{sp6}.

It is important to notice, that few methods of analytical calculations
of reactions with massive fermions are convenient for a realization on
the computer. By such, except the  above  mentioned  method  of  group
CALCUL, it is necessary to relate methods offered in
ref.~\cite{sp8},\cite{sp9}.  So in ref.\cite{sp8} the compact
formalism of an evaluation of matrix elements based on an insertion
in spinor lines of a complete  set states build  up unphysical
spinors is offered, that has allowed  to create  the  high speed
program of evaluations of Feynman amplitudes.

It is purpose of this paper to present simple method for calculating
amplitudes with using spinor products, which works for processes involving
massive fermions. The approach, that we present leads to expression for
Feynman amplitude in terms of spinor products.

\begin{center}
{\large {\bf CALCUL spinor techniques for massless fermions.} }
\end{center}

Let the four-vectors
\[
n_0=(1,0,0,0)~,n_1=(0,1,0,0),~n_2=(0,0,1,0),~n_3=(0,0,0,1)
\]
are the basis orts in Minkowski space. By using this vectors we can obtain
light-like vectors
\[
b_0=n_0-n_3,b_3=n_0+n_3,b_\lambda =n_1+i\hskip 2pt \lambda n_2,~~\lambda
=\pm 1
\]
with following properties:
\[
b_0.b_{\lambda}=0,~b_3.b_\lambda=0,~b_0.b_3=2,~b_{+}.b_{-}=-2.
\]
Next we define {\it "basic spinor"} $U_\lambda \left( b_0 \right)$ up to its
complex phase by specifying the corresponding projection operator and phase
condition:
\begin{equation}
U_\lambda \left( b_0\right) \overline{U}_\lambda \left( b_0\right) =\omega
_\lambda \not\!{b}_0,  \label{1}
\end{equation}
\begin{equation}
\frac \lambda 2\not\!{b}_\lambda U_{-\lambda }\left( b_0\right) =U_\lambda
\left( b_0\right)  \label{2}
\end{equation}
with matrix $\omega _{\lambda} = 1/2 \hskip 1pt \left( 1+\lambda
\gamma _5\right).$

The CALCUL spinor techniques for calculating processes with massless
external fermions involves the following operations.

1 step. From (\ref{1})-(\ref{2}) we construct spinors of fermions (or
antifermions) for any light-like momentum $p$ as follows
\begin{equation}
U_\lambda \left( p\right) = \not\!{p} \hskip 2pt U_{-\lambda }\left(
b_0\right) / \sqrt{2 \hskip 2pt p.b_0}.  \label{3}
\end{equation}

2 step. By using so-called 'spinor' Chisholm identity
\[
\gamma _\mu \left\{ \overline{U}_\lambda \left( p\right) \gamma _\mu
U_\lambda \left( k\right) \right\} =
\]
\begin{equation}
=2\hskip 2ptU_\lambda \left( k\right) \overline{U}_\lambda \left(
p\right) +2 \hskip 2ptU_{-\lambda }\left( p\right)
\overline{U}_{-\lambda }\left( k\right)   \label{4} \end{equation}
and equation for any four-vector $p$ with $p^2=0$
\begin{equation}
\not\!{p}=\sum_\lambda U_\lambda \left( p\right)
\overline{U}_\lambda \left( p\right) ,  \label{5} \end{equation}
we can reduce many amplitudes to expressions involving only spinor
products $ \overline{U}_\lambda \left( p\right) $ and $U_{-\lambda
}\left( k\right) $ with for vectors particles interacting in
process. The remaining possible spinor products are vanishing due to
helicity conversation. Spinor products for antifermions with
$V_\lambda \left( p\right) $ are always referred to their fermion
counterpart $U_{-\lambda }\left( p\right) $ through equation $
V_\lambda \left( p\right) =U_{-\lambda }\left( p\right) $.

Two nonzero products
\[
s\left( p,k\right) =\overline{U}_{+}\left( p\right) U_{-}\left( k\right)
=-s\left( k,p\right),
\]
\begin{equation}
t\left( p,k\right) =\overline{U}_{-}\left( p\right) U_{+}\left( k\right)
=s^{*}\left( k,p\right)  \label{6}
\end{equation}
due to equations (\ref{1})-(\ref{2}) reduce to trace
\[
\overline{U}_\lambda \left( p\right) U_{-\lambda }\left( k\right) =\lambda
/\left( \sqrt{ b_0 .p}\sqrt{b_0 .k}\right) \frac 14 \hskip 2pt Tr\left(
\omega _{-\lambda }\not\!{b}_0 \hskip 2pt \not\!{p} \hskip 2pt \not\!{k}
\hskip 2pt \not\!{b}_\lambda \right) =
\]
\begin{equation}  \label{7}
\lambda /\left( 2 \sqrt{b_0 .p}\sqrt{b_0 .k}\right) \left[p.b_0 \hskip 2pt k
.b_\lambda -k.b_0 \hskip 2pt p.b_\lambda -i\epsilon _{\mu \nu \rho \sigma
}b_0^\mu \hskip 2pt b_\lambda ^\nu \hskip 2pt p^\rho \hskip 2pt k^\sigma
\right].
\end{equation}
For light-like vectors $p =\left(p_0,p_x,p_y,p_z\right)$ and $k
=\left(k_0,k_x,k_y,k_z\right)$ we can obtained
\begin{equation}
s\left( p,k\right) =\sqrt{p_{-}k_{+}}\exp \left( i\varphi _p\right) -\sqrt{
p_{+}k_{-}}\exp \left( i\varphi _k\right)  \label{8}
\end{equation}
with $p_{\pm }=p_0\pm p_z,p_x\pm ip_y=\sqrt{p_{+}p_{-}} \exp \left( \pm
i\varphi_p\right)$. Spinor product (\ref{8}) of two four-vectors is not
complicated than simple scalar Minkowski product $p.k$.

Generalizations of CALCUL approach to the massive fermion case exist
~\cite{sp6}-\cite{sp7}, but the results are cumbersome: the method
requires the introduction of extra light-like momenta and expression
of the spinor for massive fermion as sum over spinors with extra
momenta.  We will be below, that the approach presented here, it is
possible to write amplitude in terms of spinor products for any
polarization vector of the massive fermions.

\begin{center}
{\large {\bf Spinor techniques for massive fermions.} }
\end{center}

Our procedure will be similar to that of the above-listed section.
Considering bispinors, which relating with "basic spinor" (\ref{2})
\begin{equation}
U_\lambda \left( p,s_p \right) =\frac{\tau _u^\lambda \left( p,s_p \right) }{
\sqrt{ b_0.\left( p+m_p s_p\right) }}U_{-\lambda }\left( b_0\right) ,
\label{9}
\end{equation}
\begin{equation}
V_\lambda \left( p,s_p\right) =\frac{ \tau _v^\lambda \left( p,s_p\right) } {
\sqrt{ b_0.\left( p+m_p s_p\right) }}U_\lambda \left( b_0\right),
\label{10} \end{equation}
where the projection operators $\tau
_u^\lambda \left( p,s_p\right), \tau _u^\lambda \left( p,s_p\right)$
are \begin{equation} \tau _u^\lambda \left( p,s_p\right) =\frac
1{2}\left( \not\!{p}+m_p\right) \left( 1+\lambda \hskip 2pt \gamma
_5\hskip 2pt \not\!{s_p}\right), \label{11} \end{equation}
\begin{equation}
\tau _v^\lambda \left( p,s_p\right) =\frac 1{2}\left( \not\!{p}-m_p\right)
\left( 1+\lambda \gamma _5\hskip 2pt \not\!{s_p}\right).  \label{12}
\end{equation}
We can get that
\[
\not\!{p} \hskip 2pt U_\lambda \left( p,s_p\right) = m_p \hskip 2pt
U_{\lambda }\left( p,s_p\right) , \not\!{p} \hskip 2pt V_\lambda \left(
p,s_p\right) = - m_p \hskip 2pt V_{\lambda }\left( p,s_p\right) , \]
\begin{equation}
\gamma_5 \not\!{s_p} \hskip 2pt U_\lambda \left( p,s_p\right) =
\lambda \hskip 2pt U_{\lambda }\left( p,s_p\right) , \gamma_5
\not\!{s_p} \hskip 2pt V_\lambda \left( p,s_p\right) = \lambda
\hskip 2pt V_{\lambda }\left( p,s_p\right) \end{equation}
i.e. the
bispinors $U_\lambda \left( p,s_p\right)$ and $V_\lambda \left(
p,s_p\right) $ satisfies Dirac equation and spin conditions for
massive fermion and antifermion. We also can get, that bispinors of
fermion and antifermions (\ref{9})-(\ref{10}) have relationships:
\begin{equation}
V_\lambda \left( p,s_p\right) =-\lambda \gamma _5U_{-\lambda }\left( p,s_p\right),
\overline{V}_\lambda \left( p,s_p\right) =\overline{U}_{-\lambda }\left(
p,s_p\right) \lambda \hskip 2pt \gamma_5.  \label{14}
\end{equation}

Let's analyze how many there are spinor products for massive fermions?
Obviously, that technically situation with massive fermions more difficult
on a comparison with massless. In the most general case for calculations of
matrix elements are necessary sixteen spinor products in difference
from second in massless version. And consequently, on the first
sight, it seems, that the use of a reduction of a matrix element of
process to spinor products is not absolutely convenient. However,
here it is possible to achieve essential simplification. At first,
all spinor products are not linearly independent.  Using ratios
(\ref{14}), that only eight are linearly independent.  We define
{\it "basic spinor products" } for massive fermions as
\begin{equation} \begin{array}{c}
\overline{U}_\lambda \left( p,s_p\right) U_{-\lambda }\left( k,s_k\right) ,
\\
\overline{U}_\lambda \left( p,s_p\right) U_\lambda \left( k,s_k\right) , \\
\overline{V}_\lambda \left( p,s_p\right) U_\lambda \left( k,s_k\right) , \\
\overline{V}_\lambda \left( p,s_p\right) U_{-\lambda }\left( k,s_k\right) .
\end{array}
\label{13}
\end{equation}

Secondly, basic spinor products can be calculated with the help of second
functions. For it is used, that $\gamma _5$ for massless bispinor is the
operator of a spin i.e.
\begin{equation}
\gamma _5U_\lambda \left( b_0\right) =\lambda U_\lambda \left( b_0\right) .
\label{fer1}
\end{equation}

Then definition of bispinors (\ref{9})-(\ref{10}) is possible to rewrite
as:
\begin{equation}
U_\lambda \left( p,s_p \right) = \frac{\chi\left( p,s_p,+1 \right) }{2 \sqrt{
b_0.\left( p+m_p s_p\right) }}U_{-\lambda }\left( b_0\right) ,
\label{fer2} \end{equation} \begin{equation} V_\lambda \left(
p,s_p\right) =\frac{ \chi \left( p,s_p,-1 \right) } {2 \sqrt{
b_0.\left( p+m_p s_p\right) }}U_\lambda \left(b_0\right)
\label{fer3} \end{equation}
with functions
\begin{equation} \chi
\left( p,s_p,a \right) = \left( \not\!{p}+a \hskip 1pt m_p \right)
\left( 1+a \hskip 1pt \not\!{s_p} \right).  \label{fer4}
\end{equation}

Let's introduce the following functions:
\[
s \chi \left( p,k,s_p,s_k,\lambda,a,b \right) = \lambda/ \left( 2
\sqrt{ b_0. \left( p+m_p \hskip 2pt s_p \right)} \sqrt{ b_0. \left(
k+m_k \hskip 2pt s_k \right) } \right) \]
\begin{equation} \frac 14
Tr \left( \omega_{-\lambda }\hskip 2pt \not\!{b}_0 \hskip 2pt
\chi^{\dagger} \left( p,s_p,a \right) \chi \left( k,s_k,b \right) \hskip 2pt
\not\!{b}_\lambda \right),  \label{fer5}
\end{equation}
\[
w\chi \left(p,k,s_p,s_k,\lambda,a,b \right) = 1/ \left( \sqrt{ b_0.
\left( p+m_p \hskip 2pt s_p \right)} \sqrt{ b_0. \left( k+m_k \hskip
2pt s_k \right) } \right) \]
\begin{equation} \frac 14 Tr \left(
\omega_{-\lambda } \hskip 2pt \not\!{b}_0 \hskip 2pt \chi^{\dagger}
\left( p,s_p,a \right) \chi \left( k,s_k,b \right) \right).
\label{fer6}
\end{equation}
All "basic" spinor products are reduced to functions (\ref{fer5}) - (\ref
{fer6}):
\begin{equation}
\begin{array}{c}
\overline{U}_\lambda \left( p,s_p\right) U_{-\lambda }\left(
k,s_k\right)=s\chi \left(p,k,s_p,s_k,\lambda,a=+1,b=+1 \right), \\
\overline{U}_\lambda \left( p,s_p\right) U_\lambda \left( k,s_k\right)=w\chi
\left(p,k,s_p,s_k,\lambda,a=+1,b=+1 \right), \\
\overline{ V}_{\lambda} \left( p,s_p\right) U_\lambda \left(
k,s_k\right)=s\chi \left(p,k,s_p,s_k,-\lambda,a=-1,b=+1 \right), \\
\overline{V}_{\lambda} \left( p,s_p\right) U_{-\lambda
}\left(k,s_k\right)=w\chi \left(p,k,s_p,s_k,-\lambda,a=-1,b=+1 \right).
\end{array}
\label{fer7}
\end{equation}

As well as for massless of fermions, the spinor products (\ref{fer7}) can be
calculated through components of vectors $p,k,s_p,s_k$. Obviously, that the
analytical expression is more complicated on a comparison with appropriate
ratios (\ref{7}) - (\ref{8}).

Essential simplification as in function evaluation to which the "basic"
spinor products are reduced it is possible to achieve at the expense of
specific selection of polarization vectors of massive fermions.

The polarization vector $s_p$  of  a  fermion  can  be  expressed
through momentum of this fermion and any other vector of a problem.
\begin{equation}
s_p=\frac{p.q \hskip 2pt p -m^2_p \hskip 2pt
q}{m_p\sqrt{(p.q)^2-m^2_p \hskip 2pt q^2}} \label{fer7a}
\end{equation}
Here  $q$-any vector. It is easy to be convinced, that $s_p$
satisfies to standard conditions.
By selection $q=b_0$ we have a polarization  vector used
in ref.\cite{sp6}, and if $q = n_0 = (1,0,0,0)$, we shall receive
helicity state. Selecting a polarization vector $s_p $ for
a fermion (antifermion) with momentum $p $ and mass $m_p $ as
\begin{equation}
s_p = p/m_p-m_p \hskip 2pt \frac{b_0}{p.b_0}.  \label{16}
\end{equation}
let's receive, that the function $\chi \left( p,s_p,a\right) $,
included in definitions of bispinors (\ref{fer2})-(\ref{fer3})
receives a kind
\begin{equation} \chi \left( p,s_p,a\right) =2\left(
\not\!{p}+a\hskip 1pt m_p\right) , \label{fer8} \end{equation}
and the factor becomes equal $2\sqrt{2\hskip 1pt b_0.p}$.

It, queue results in sharp simplification of expressions for functions
(\ref{fer5})-(\ref{fer6}) through components of vectors. These
functions now do not depend on polarization vectors of fermions and
can be calculated with the help of following formulas:  \[ s\chi
\left( p,k,s_p,s_k,\lambda,a,b\right) \equiv s_\lambda \left(
p,k\right) =
\]
\begin{equation}
\lambda /\left( 2\sqrt{b_0.p}\sqrt{b_0.k}\right) \left[ p.b_0\hskip
2ptk.b_\lambda -k.b_0\hskip 2ptp.b_\lambda -i\lambda \epsilon _{\mu \nu \rho
\sigma }b_0^\mu \hskip 2ptb_\lambda ^\nu \hskip 2ptp^\rho \hskip 2ptk^\sigma
\right] ,  \label{fer9}
\end{equation}
\[
w\chi \left( p,k,s_p,s_k,\lambda ,a,b\right) \equiv w\left( p,k,a,b\right) =
\]
\begin{equation}
1/\left( \sqrt{b_0.p}\sqrt{b_0.k}\right) \left[ am_p\hskip 2ptk.b_0+bm_k
\hskip 2ptp.b_0\right] .  \label{fer10}
\end{equation}

The formula (\ref{fer9}) does not differ from a similar ratio for case of
massless fermions (\ref{7}) (only it is necessary to remember, that $p^2 =
m^2 _p, k^2 = m^2_k $). In addition, there is a function (\ref{fer10}). For
calculations of matrix elements with massive fermions the group
CALCUL of ~\cite{sp6} used fermions with polarization vectors
(\ref{16}), but with introduction additional for each fermions
light-like of momentums, that actually did a procedure of
evaluations is much more complex on a comparison with similar for
the massless fermions. In our case "basic" products practically
differ from similar addends in version with massless fermions, but
we not enter expansions of vectors of process on light-like vectors.
In the massless limit, our expressions easy simplifies to
spinor products $s\left( p,k\right)$ and $ t\left( p,k\right)$
(\ref{6}).
\begin{center}
{\large {\bf Chisholm indenties for massive fermions.} }
\end{center}

The important stage in transformation  of  matrix  elements  to  basis
spinor products is the use of "spinor" identities Chisholm  of  a type
(\ref {4}). For a proof of this ratio (see,\cite {sp6})  the
Chisholm identity  for trace was used:
$$
\gamma_{\mu}*Tr(\gamma_{\mu} \not\!{C}_1\ldots \not\!{C}_{2n+1})=
$$
\begin{equation}
=2 \left( \not\!{C}_1 \ldots \not\!{C}_{2n+1}
+ \not\!{C}_{2n+1}\ldots \not\!{C}_1\right) .
\label{fer11}
\end{equation}

As the expression of a type $ \overline{U}_\lambda  \left(p,s_p\right)
\gamma_\mu U_\lambda \left (k,s_k\right)$ is reduced
to trace, using the formulas (\ref {9}), \ref {10}), (\ref {fer11}) it
is possible to receive appropriate  "spinor"  identities  for  massive
fermions. In this case is present 4 (four) such "basic" of identities
(in difference from one in massless version):
\[ \gamma _\mu \left\{
\overline{U}_\lambda \left( p,s_p\right) \gamma _\mu U_\lambda
\left( k,s_k\right) \right\} = \] \[ U_\lambda \left( k,s_k\right)
\overline{U}_\lambda \left( p,s_p\right) +U_{-\lambda }\left(
p,s_p\right) \overline{U}_{-\lambda }\left( k,s_k\right) + \]
\begin{equation}
+V_{-\lambda }\left( k,s_k\right) \overline{V}_{-\lambda }\left(
p,s_p\right) +V_\lambda \left( p,s_p\right) \overline{V}_\lambda \left(
k,s_k\right) ,  \label{fer12}
\end{equation}
\[
\gamma _\mu \left\{ \overline{U}_\lambda \left( p,s_p\right) \gamma _\mu
U_{-\lambda }\left( k,s_k\right) \right\} =
\]
\[
U_{-\lambda }\left( k,s_k\right) \overline{U}_\lambda \left( p,s_p\right)
-U_{-\lambda }\left( p,s_p\right) \overline{U}_\lambda \left( k,s_k\right) +
\]
\begin{equation}
+V_\lambda \left( p,s_p\right) \overline{V}_{-\lambda }\left( k,s_k\right)
-V_\lambda \left( k,s_k\right) \overline{V}_{-\lambda }\left( p,s_p\right) ,
\label{fer13}
\end{equation}
\[
\gamma _\mu \left\{ \overline{V}_\lambda \left( p,s_p\right) \gamma _\mu
U_{-\lambda }\left( k,s_k\right) \right\} =
\]
\[
U_{-\lambda }\left( k,s_k\right) \overline{V}_\lambda \left( p,s_p\right)
+V_{-\lambda }\left( p,s_p\right) \overline{U}_\lambda \left( k,s_k\right)
\]
\begin{equation}
+V_\lambda \left( k,s_k\right) \overline{U}_{-\lambda }\left( p,s_p\right)
+U_\lambda \left( p,s_p\right) \overline{V}_{-\lambda }\left( k,s_k\right) ,
\label{fer14}
\end{equation}
\[
\gamma _\mu \left\{ \overline{V}_\lambda \left( p,s_p\right) \gamma _\mu
U_\lambda \left( k,s_k\right) \right\} =
\]
\[
U_\lambda \left( k,s_k\right) \overline{V}_\lambda \left( p,s_p\right)
+U_\lambda \left( p,s_p\right) \overline{V}_\lambda \left( k,s_k\right)
\]
\begin{equation}
-V_{-\lambda }\left( k,s_k\right) \overline{U}_{-\lambda }\left(
p,s_p\right) -V_{-\lambda }\left( p,s_p\right) \overline{U}_{-\lambda
}\left( k,s_k\right) .  \label{fer15}
\end{equation}

The remaining possible combinations are reduced to  above-stated  with
the help of ratioes (\ref {14}). In case massless fermions of  a ratio
(\ref{fer12}),  (\ref{fer14})  coincide  one  another  (to  within
replacement $\lambda \to -\lambda$ and pass in to (\ref  {4}),
and in  identities  (\ref{fer13}),  (\ref{fer15})  the  right
member addresses in a zero, owing to a conservation helicity law.
Using (\ref {fer12})-(\ref{fer15}),  and  also  Dirac equation and
condition of a completeness for bispinors  we  can  record
analytical expression of a matrix element in terms of spinor
products.

\end{document}